\documentclass{aa}
\usepackage{graphics}

\def\ltapprox{\raise 2pt \hbox {$<$} \kern-1.1em \lower 5pt \hbox {$\approx$}}
\def\ltsim{\raise 2pt \hbox {$<$} \kern-1.1em \lower 4pt \hbox {$\sim$}}
\def\gtsim{\raise 2pt \hbox {$>$} \kern-1.1em \lower 4pt \hbox {$\sim$}}

\begin{document}

\title{Modeling the interaction between ICM and relativistic plasma in 
cooling flows: the case of the Perseus Cluster}

\author{M. Gitti\inst{1,2}
\and G. Brunetti\inst{2}
\and G. Setti\inst{1,2}} 

\offprints{Myriam Gitti, c/o Istituto di Radioastronomia del CNR, via Gobetti
101, I--40129 Bologna, Italy; mgitti@ira.cnr.it}

\institute{Dipartimento di Astronomia, 
via Ranzani 1, I--40127 Bologna, Italy
\and
Istituto di Radioastronomia del CNR,
via Gobetti 101, I--40129 Bologna, Italy 
}

\authorrunning{Gitti et al.}
\titlerunning{Interaction between ICM and relativistic plasma in cooling flows}
\date{Received 23 November 2001 / Accepted 14 February 2002}

\abstract{
We argue that the presence of diffuse synchrotron emission forming the 
so-called \textit{radio mini-halos} in some cooling flow clusters can be 
explained by reacceleration of relativistic electrons,
the necessary energetics being supplied by the cooling flows themselves.
In particular, the reacceleration due to MHD turbulence has the correct 
radial dependence on the parameters to naturally 
balance the radiative losses.
As an application we show that the main properties of the radio mini-halo 
in the Perseus cluster (brightness profile, total radio spectrum and 
radial spectral steepening) can be accounted for by the synchrotron radiation
from relic relativistic electrons in the cluster, which are efficiently
reaccelerated by MHD turbulence \textit{via} 
Fermi--like processes due to the compression of the cluster magnetic 
field in the cooling flow region.
Since the presence of an observable radio mini-halo in a cooling flow region 
critically depends on the combination of several physical parameters, 
we suggest that the rarity of radio mini-halos found in cooling flow clusters 
is due to the fact that the physical conditions of the ICM are intermediate 
between those which lead to the formation of extended radio halos and those 
holding in cooling flows without radio halos.
The basic results of our model remain unchanged even if the cooling flow is
stopped somewhere in the innermost region of the cluster. 
\keywords{acceleration of particles -- 
radiation mechanisms: non-thermal --
cooling flows --
galaxies: clusters: general -- 
galaxies: clusters: individual: Perseus (A426) }
}

\maketitle

\section{Introduction}
\label{intro}

The intracluster medium (ICM) consists of a hot gas emitting thermal 
X--ray, of large scale magnetic fields and of relativistic plasma. 
It is well known that due to synchrotron and inverse Compton (IC) losses, 
the typical ageing time--scale of the relativistic electrons in the ICM is 
relatively short ($10^7 \div 10^8$ yr) so that the electrons should already 
have lost most of their energy.
Nevertheless, diffuse synchrotron emission from clusters of galaxies has
been detected by radio observations in a number of cases (e.g. Feretti 2000). 
Thus one can suppose that relativistic electrons may be reaccelerated by 
some mechanisms acting with an efficiency comparable to the energy
loss processes (e.g. Petrosian 2001 and references therein) 
or, alternatively, one should turn to secondary electrons models 
(Dennison 1980, Blasi \& Colafrancesco 1999, 
Dolag \& En{\ss}lin 2000). 
\\
Large--scale radio halos in clusters of galaxies appear as diffuse radio 
sources of low surface brightness, steep radio spectrum and typical  
Mpc--size, not directly associated with the galaxies but rather diffused into 
the ICM at the center of the clusters (Feretti \& Giovannini 1996 and 
references therein).
Liang et al. (2000) have shown the presence of a correlation between radio 
power and cluster temperature for radio halos with good radio and X--ray 
data available. This correlation, together with the morphological similarity 
of thermal X--rays and radio emission (Govoni et al. 2001),
suggests a physical relationship between the properties of thermal ICM 
and the presence of radio halos.
It is also well known that there exists an anti--correlation between 
the presence of a cooling flow and of a radio halo at 
the cluster center: radio halos are rarely found in clusters of galaxies 
with cooling flow and, when they are, they appear quite different from 
canonical halos such as that in the Coma cluster (Tribble 1993).
However, there are a few cooling flow clusters where the relativistic 
particles can be traced out quite far, forming what is called a 
``\textit{mini--halo}'': a radio source smaller in extent, of low surface 
brightness and steep spectral index, around a powerful radiogalaxy  at the 
center of a cluster (e.g. Virgo: Owen et al. 2000; Perseus: Burns et al. 1992;
PKS 0745 - 191: Baum \& O'Dea 1991). 
\\
On the basis of the above considerations, it is clear that 
the persistent diffuse radio emission, and hence the need for reacceleration 
of the relativistic plasma, may be found in the interaction between  
the thermal and the relativistic component in the ICM. Indeed,
Brunetti et al. (2001) have proposed that the relativistic electrons injected
in the cluster volume by AGN and/or galactic winds in the course of the
cluster evolution may form a relic electron population and that an extended 
radio halo is originated if the relic, but still relativistic, electrons are 
reaccelerated to higher energies by shocks and/or turbulence in the ICM. 
In this model, as suggested by observations, the merger events may supply 
the necessary energy for the reacceleration.
In this paper we argue that a similar model holds for the relativistic 
electrons in the mini--halos too, and that the necessary energetics is 
supplied by the cooling flows, as suggested by Tribble (1993) and Sijbring 
(1993). Unlike the case of the extended radio halos, the energy can not arise 
from the merger events, since observations in general 
show no mergers acting in clusters with cooling flows (Edge et al. 1992).
\\ 
The main aim of the present work is to investigate whether the reacceleration
of relic electrons and the increased intensity of the cluster magnetic field
due to the compression in the turbulent cooling flow can produce 
radio mini--halos.       
The model expectations are then compared with the
observational properties of the radio emission from the mini--halo in 
the Perseus cluster, where a massive cooling flow is present (Ettori et al.
1998).

~\\
In Section 2 we consider the effect of a cooling flow on the intracluster 
magnetic field and calculate the electron energy distribution
subject to reacceleration and losses. In Section 3 we present the results 
concerning the radio properties of the radio mini--halo in the 
Perseus cluster.
Our conclusions are given in Section 4.
\\
$\mbox{H}_0 = 50 \mbox{ km s}^{-1} \mbox{ Mpc}^{-1}$ 
is assumed in this paper and, where not specified, all the formulae are in 
cgs system. 

\section{The models for electron reacceleration}
\label{model}

\subsection{Magnetic fields and relativistic electrons in cooling flows}
\label{magnetic}

X-ray and optical observations indicate that large amounts of gas are cooling
and flowing into the centers of clusters of galaxies;
the observations and theory of ``\textit{cooling flows}'' are
reviewed in Fabian (1994).
For thermal bremsstrahlung models, the decrease of the temperature with the 
distance $r$ from the center is $T(r) \propto r^{1.2}$, whereas the gas 
density increases towards the center with the same behaviour 
(Fabian et al. 1984). 
The compression of the gas is expected
to produce a sensible increase of the strength of the frozen--in 
intracluster magnetic field. 
Depending on the physical conditions prevailing in a cooling flow, 
in particular on the values of the turbulent velocity $v_T$ and
of the mean inflow velocity $v_F$, there are two possibilities: 
\begin{itemize}
\item if $v_T \ll v_F$, the inward cooling flow is homogeneous and 
the turbulence does not distort the field geometry during the infall. 
In this case we can follow Soker \& Sarazin's model (1990).
They have presented numerical solutions showing that 
the frozen--in magnetic field is greatly increased by the compression of the 
ICM and its lines become predominantly radial as the gas flows 
inward. Under the assumption that the field is not dynamically important, 
one has that
\begin{equation} \label{camposoker}
B(x) = \frac{B_c}{\sqrt{3}} \; x^{-2} \left(1 + 2 x^{\frac{18}{5}}\right)
^{\frac{1}{2}} \sim x^{-2}
\end{equation}
where $r_c$ is the cooling radius, $x=r/r_c$ is the adimensional 
distance from the center
and $B_c$ is the field strength at $r_c$.\\ 
\begin{figure}
\includegraphics{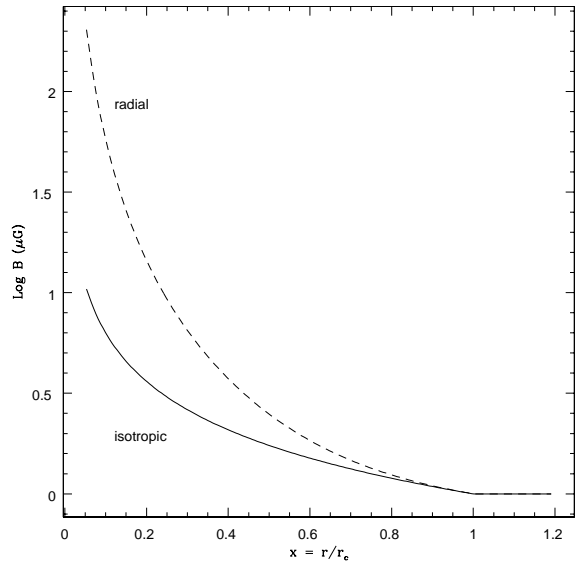}
\vspace{8 cm}
\caption{The calculated field growth in the cooling flow region 
for Soker \& Sarazin's model (radial field, dashed line) and Tribble's model 
(isotropic field, solid line). We have assumed $r_c = 210$ kpc and 
$B_c = 1 \, \mu$G.} 
\label{campops}
\end{figure}
\item if $v_T \gg v_F$, as pointed out by Tribble (1993), an individual element
of gas in the flow is in constant motion relative to the mean flow and 
its mean position falls slowly inwards. 
The overall effect is an isotropic compression of the field. 
In this case the variation of the field with the gas density 
$n$ within the central cooling region is 
$B \propto n^\frac{2}{3}$, so that being $n(x) \propto x^{-1.2}$, 
it results
\begin{equation} \label{campotribble}
B(x) = B_c \; x^{-0.8}
\end{equation}
\end{itemize}
~\\
The resulting intensity profile of the magnetic field in these two different 
cases is plotted in Fig. \ref{campops}.
In both cases the field strength grows towards the center,
but with quite different behaviours.
In Soker \& Sarazin's model the field strength grows faster than in Tribble's 
one and in the central region the field reaches values which are 
at least one order of magnitude larger than those reached by the isotropic 
compression.
It should be mentioned that intense fields of the order of some tens of 
$\mu$G have been derived by 
detections of extremely high Faraday rotation measures throughout 
radio galaxies in the center of cooling flow clusters 
(Taylor \& Perley 1993; Ge \& Owen 1993; Taylor et. al. 1990; 
Perley \& Taylor 1991). This has important consequences on the evolution of 
the relativistic electron energy and spectrum. 
Indeed, the radiative life--time of an ensemble of relativistic electrons 
losing energy by synchrotron emission and IC scattering off the cosmic 
microwave background photons (CMB) is given by
\begin{equation}  
\label{tau.sinic}
\tau_{sin+IC} = \frac{16.2}{\left[B_{\perp}^2 + \frac{2}{3}
B_{CMB}^2 \right] \, \gamma} \, \, \mbox{ yr} 
\end{equation}
where $\gamma$ is the Lorentz factor and
$B_{CMB} = 3.18 (1+z)^2 \; \mu$G denotes the magnitude of the 
magnetic field equivalent to the CMB from the standpoint of the losses.
Without any reacceleration mechanism at work, 
the emission of these electrons
in the radio band should not be observable for more than $\sim 10^8$ yr. 
This short lifetime contrasts with the diffuse radio emission observed, 
for example, in the Perseus cluster: 
whatever the model for the origin and presence of the relativistic electrons 
in the mini--halo region is, this characteristic lifetime
is both shorter than the crossing time 
(both assuming the Alfv\'en velocity or the diffusion
coefficient as in Sarazin 2001),
and shorter than the cooling time (if one assumes that the 
radio--emitting region is filled with electrons carried by the cooling flow). 
Hence, it seems plausible that the electrons have been reaccelerated.
\\
A reaccelerated relic electron population in cluster radio halos would 
naturally produce radio synchrotron spectra which steepen with increasing 
distance from the center (as observed in the Coma radio halo), whereas 
it can not be produced in the case of models involving a secondary electron 
population (Brunetti et al. 2001). 
Here we stress that the radio spectral index map of the mini--halo in the 
Perseus cluster shows a progressive steepening of the spectral index with 
increasing distance (Sijbring 1993), thus providing an additional indication
for the presence of an effective mechanism for electron 
reacceleration operating in the ICM in the cooling flow region.
 
\subsection{Evolution of electron energy spectrum into the cooling flow}
\label{evolution}

In our model we assume that the relativistic electrons are continuosly 
reaccelerated by \textit{Fermi type} mechanisms. Given an initial
monoenergetic electron distribution, the electrons are \textit{systematically} 
accelerated with an efficiency $\alpha_2$ and, at the same time,
their energy is \textit{stocastically} changed with an efficiency $\alpha_1$.
These two combined effects result in a relatively wide distribution
of the energy of the electrons around a mean value determined by systematic 
processes. 
In particular, it is well known that MHD turbulence can efficiently 
accelerate electrons \textit{via} Fermi--like processes.
In order to allow relativistic electrons to still be emitting in the radio
band, the reacceleration mechanism has to act with an efficiency comparable 
to the energy loss processes.
Due to the competition of these two effects, at any distance $x$ 
the time evolution of the energy of a relativistic electron is given by:

\begin{equation} \label{dgamma}
\frac{d\gamma(x)}{dt} = -\beta(x) \gamma^2(x) + \alpha_2(x) \gamma(x) - 
\chi(x)
\end{equation}
where $\beta(x)$ is the coefficient of synchrotron and IC losses,
$\alpha_2(x)$ the acceleration coefficient and $\chi(x)$ the Coulomb 
losses term.
The Coulomb losses depend on the ICM density as (e.g Sarazin 1999): 
\begin{equation} \label{coulomb}
\chi(x) \mathbf{ \sim }  1.4 \times 10^{-12} n(x) \; \; \mbox{ s}^{-1}
\end{equation}
where $n(x) = n_c \; x^{-1.2}$,
$n_c$ being the electron density at the cooling radius.
\\
The coefficient of the radiative losses is:
\begin{equation} \label{beta}
\beta(\vartheta,x) = 1.9 \times 10^{-9} \left[B^2(x) \sin^2 \vartheta +
\frac{2}{3} B_{CMB}^2 \right] \; \; \mbox{ s}^{-1}
\end{equation}
where $\vartheta$ is the pitch angle and where we have neglected the
dependence on the redshift, not important in our model.
If the pitch angle scattering is an efficient process, then the electrons 
will be continuosly isotropized and the term $B^2(x) \sin^2 \vartheta$ 
should be replaced by $2/3 B^2(x)$, where $B(x)$ is obtained from 
Eq. \ref{camposoker} or \ref{campotribble}.
\\
According to the usual Fermi acceleration theory (e.g. Melrose 1980), it is
$\alpha_2 = 2 \alpha_1$ and the coefficient of systematic Fermi acceleration 
due to MHD turbulence is given by:
\begin{equation} \label{coeffermi}
\alpha_2(x) \sim \frac{v_A^2(x)}{l(x) \: c} \left(
\frac{\delta B(x)}{B(x)}\right)^2
\end{equation}
where $v_A(x) \sim 2.2\times 
10^{11} \: n_c^{-0.5} \; x^{0.6} \; B(x) \; \; \mbox{ cm s}^{-1}$,
and $l(x)$ is the characteristic distance between two subsequent peaks of 
magnetohydrodynamic turbulence.\\
We assume the existence of an energetic turbulence such that 
$\delta B(x) / B(x) \sim 1$ constant through the whole cooling flow region.
Since the characteristic time of radiative losses depend on $\gamma$, 
while that of Fermi acceleration is independent of $\gamma$,
the losses dominate the time evolution of the electrons for energies higher 
than $\gamma_b$, the \textit{break energy}. 
It is worth noticing that MHD turbulence can efficiently accelerate 
electrons \textit{via} Fermi--like processes when their velocity parallel to 
the magnetic field is $v_{\|} > (m_p/(m_e \mu))^{1/2} \times v_A$, 
where $v_{\|} = v \mu$ (Eilek \& Hughes 1991; Hamilton \& Petrosian 1992). 
The fact that in our model we deal with a seed population of relativistic 
electrons guarantees the efficiency of this process.
Under the simple assumption that the total number of turbulent 
elements is conserved during the infall, $l(x)$ is roughly proportional to 
$n(x)^{-\frac{1}{3}}$, so that $l(x) = l_c \; x^{0.4}$, $l_c$ denoting
the characteristic length at the cooling radius.
The underlying assumption is that the turbulence does not decay
on a time scale shorter than the acceleration time scale.
\\
By solving Eq. \ref{dgamma} at any time $t$, 
one can calculate the break energy $\gamma_b(t)$, corresponding to the energy 
at the time $t$ of the electrons 
having a near infinite energy at the initial time $t=0$. 
By omitting for simplicity of notation the dependence on the  
distance from the cluster center, it results (Brunetti et al. 2001):

\begin{equation} 
\label{gammabkt}
\gamma_b(t) = \frac{1}{2 \beta} \left(\alpha_2 + \frac{\sqrt{{\alpha_2}^2 -
4\beta\chi}}{\tanh \left(\frac{t}{2} \sqrt{{\alpha_2}^2 - 4\beta\chi}\right)}
\right)
\end{equation}
The break energy of the stationary spectrum, which corresponds to the energy 
at which the losses are balanced by the reacceleration, is obtained by 
evaluating Eq. \ref{gammabkt} in the limit $t \rightarrow \infty$:
    
\begin{equation} 
\label{gammabk}
\gamma_b = \frac{1}{2 \beta} \left(\alpha_2 + \sqrt{{\alpha_2}^2 - 4\beta\chi}
\right)
\end{equation}
Because of the distance dependence of the coefficients,
the break energy in the electron spectrum is not the same throughout 
the whole cooling flow region, but it (weakly) depends on the 
radial distance.
\\
The time evolution of the electron energy distribution is obtained
by solving the continuity equation (Kardashev 1962) taking into account the 
acceleration by the Fermi mechanism and the losses.
\\
Since the acceleration time scale (Eq. \ref{coeffermi}) 
\begin{equation}
\label{tauacc}
\tau_{acc} = \alpha_2^{-1}
\sim 6 \times 10^{7} \; l (\mbox{pc}) 
\; \left(\frac{n}{10^{-3}} \right) B^{-2} (\mu\mbox{G}) \; \; 
\mbox{ yr}
\end{equation}
is much shorter 
(e.g. $\tau_{acc} \sim 8 \times 10^8$ yr for the values of parameters 
obtained for the Perseus cluster at the cooling radius, 
Section \ref{results}) 
than the time it takes for a 
volume element to be convected by the flow from $r_c$ into the
center, we can solve the continuity equation under 
stationary conditions, obtaining (for $\gamma \gg 1$)
\begin{equation} 
\label{spettroel}
N(\gamma) = N(\gamma_b) \exp(-\Phi) \cdot
\left(\frac{\gamma}{\gamma_b}\right)^2 \exp 
\left[f\left(\frac{\gamma}{\gamma_b}\right)\right]
\end{equation}
where $N(\gamma_b)$ is the number of electrons at $\gamma_b$,\\
$\Phi=2-\frac{4}{1+\frac{\chi}{\beta {\gamma_b}^2}}$ and
\begin{equation}
f\left(\frac{\gamma}{\gamma_b}\right) = \frac{\frac{2 \chi}{{\gamma_b}^2} - 
2 \beta {\left(\frac{\gamma}{\gamma_b}\right)}^2}{\left(\beta+ \frac{\chi}
{{\gamma_b}^2}\right) \left(\frac{\gamma}{\gamma_b}\right)}
\end{equation}
Since in our case it is $\chi/(\beta \gamma_b^2) \ll 1$ and 
$\chi/(\beta \gamma_b^2) \ll (\gamma/\gamma_b)^2$, 
Eq. \ref{spettroel} can be approximated as
$N(\gamma) \propto \left(\gamma/\gamma_b \right)^2 
\exp\left(-2 \gamma/\gamma_b \right)$, 
in agreement with the result of Borovsky and Eilek's (1986), that indeed have 
not considered the Coulomb losses.
\\
Note that the spectral distribution of the electrons 
$N(\gamma)$ (Eq. \ref{spettroel}) depends on $\gamma/\gamma_b$ only, 
while the values of the coefficients simply produce a ``normalisation'' 
of the spectrum which depends on the radial distance.
\\
In principle, in order to correctly derive the energy density of the 
relativistic electrons as a function of $x$ one should solve the spatial 
diffusion equation for the relativistic electrons starting from some 
initial distribution of the number density and spectrum of the 
electrons which, however, are basically unknown.
As a consequence, in this paper we prefer to parameterize the electron 
energy density, essentially peaked at $\gamma \sim \gamma_b$, as:

\begin{equation}
\label{norms}
m_e c^2 \int_1^{\infty}N(\gamma)\gamma d\gamma \approx m_e c^2 N(\gamma_b)
{\gamma_b}^2 \propto x^{-s}
\end{equation}
where $s$ is a free parameter which will be constrained and discussed in  
Section \ref{results}.

\subsection{The synchrotron spectrum}
\label{spectrum}

The synchrotron emissivity of a population of relativistic electrons 
with energy distribution per unit solid angle $N(\gamma, \Omega)$ is given by: 
\begin{equation}
J_{\nu} = \int_{\Omega} \int_1^{\infty} p(\nu,\vartheta)
N(\gamma, \Omega) d\gamma d\Omega
\end{equation}     
where $p(\nu,\theta) = \frac{\sqrt{3} e^3 B \sin \vartheta}{m_e c^2} F
\left(\frac{\nu}{\nu_c}\right)$ is the emitted power per unit frequency 
and per unit solid angle from each electron, 
$F\left(\frac{\nu}{\nu_c}\right)$ being the Kernel function and 
$\nu_c = \frac{3}{4 \pi} \frac{e B \sin \vartheta }{m_e c} \gamma^2$ 
the critical frequency (Ginzburg \& Syrovatskii 1965).
Under the assumption of an isotropic distribution of electron  
momenta and of the magnetic field lines we can assume the synchrotron 
emissivity to be isotropic, if averaged over a sufficiently large volume.
\\
With the energy distribution of the relativistic electrons given by 
Eq. \ref{spettroel}, the synchrotron emissivity per unit solid angle is:
\begin{eqnarray} \label{spettro}
J_{\nu}(x) &=& K(x) \cdot
\int_0^{\frac{\pi}{2}} \int_1^{\infty} \sin^2 \vartheta \cdot F\left(\frac{\nu}
{\nu_b} \frac{{\gamma_b}^2}{\gamma^2 \sin \vartheta}\right) \cdot \gamma^2
\cdot \nonumber \\
~ & \cdot & \exp\left[f\left(\frac{\gamma}{\gamma_b}\right)\right] 
d\vartheta d\gamma 
\end{eqnarray}
where 
\begin{equation} \label{kspettro}
K(x) = \frac{\sqrt{3} e^3}{m_e c^2} \cdot B(x) 
\cdot N(\gamma_b)_c \; \frac{{\gamma_{b,c}}^2}{{\gamma_b}^4} \cdot \exp(-\Phi) 
\cdot x^{-s}
\end{equation}
$\nu_b$ being the critical frequency for $\gamma=\gamma_b \;
(\vartheta=90^{\circ})$.
The shape of the synchrotron spectrum is plotted in Fig. \ref{spettros}:
\begin{figure}[h]
\includegraphics{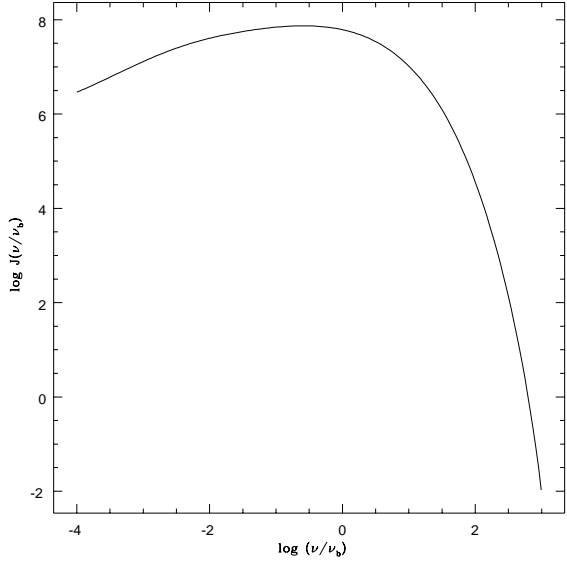}
\vspace{8 cm}
\caption{The predicted synchrotron spectrum is shown in arbitrary units.} 
\label{spettros}
\end{figure}
at low frequencies the synchrotron spectrum is approximately proportional to
$\nu^ {- 1/3}$, as in the case of the emission by a single electron, 
it becomes roughly flat just below the break frequency
(contributed by the electrons at the break energy $\gamma_b$) and then it 
falls down exponentially.

\section{Model results for the Perseus cluster }
\label{results}

\begin{figure}[h]
\includegraphics{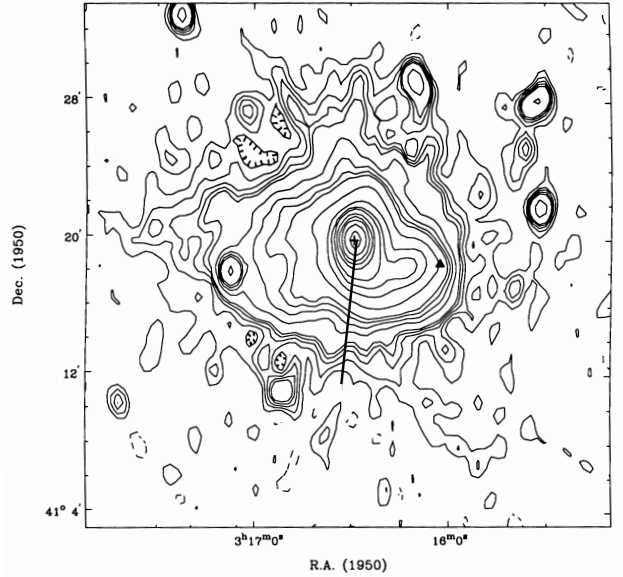}
\vspace{8 cm}
\caption{92 cm map of 3C 84 at a resolution of $51'' \times 77''$ 
(Sijbring 1993). The contour
levels are -4 (dashed), 4, 8, 12, 16, 20, 40, 60, 80, 100, 200, 400, 600, 800,
1500, 2500, 4000, 8000, 16000 mJy/beam. The r.m.s. noise is 1.4 mJy/beam.
The cross indicates the position of NGC1275 and the solid line represents the
direction we have considered.} 
\label{mappa}
\end{figure}

\begin{figure}[h]
\includegraphics{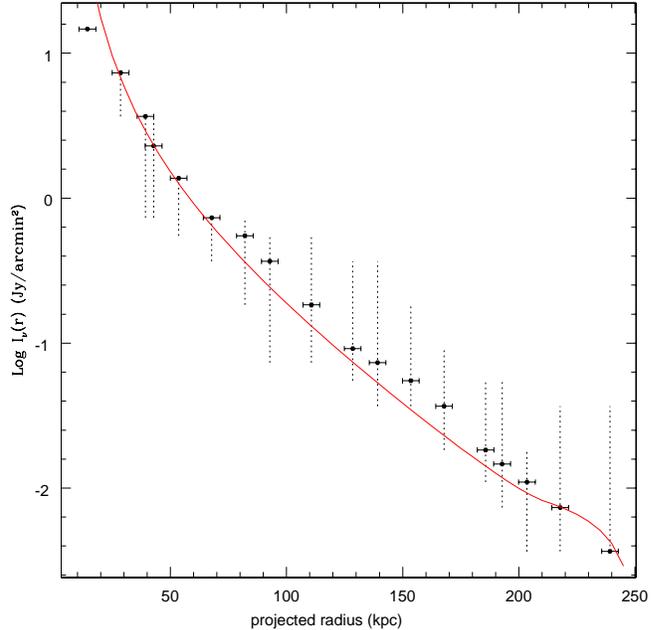}
\vspace{8cm}
\caption{Fit to the surface brightness profile obtained in the case of 
isotropic field with the following set of parameters: 
$B_c=1.2 \mu\mbox{G}$, $\mathbf{l_c=25 \mbox{ pc}}$, 
$s=2.1$.  
Vertical errorbars represent the deviations from the spherical symmetry 
of the diffuse radio emission in other directions in the cluster with respect 
to the one we have considered.}
\label{fit}
\end{figure}

\begin{figure}[h]
\includegraphics{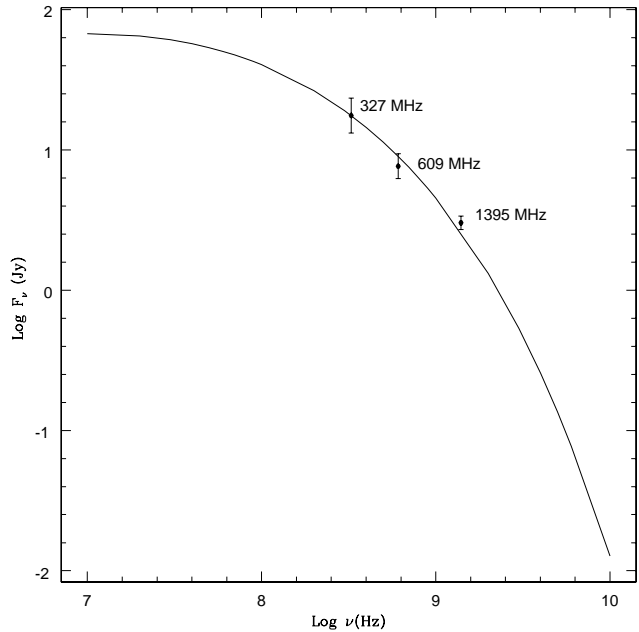}
\vspace{8cm}
\caption{\textit{Solid line}: fit to the total spectrum obtained 
with the same set of parameters of Fig. \ref{fit}. 
The radio data are taken from Sijbring (1993).}
\label{flussi}
\end{figure}

\begin{figure}[h]
\includegraphics{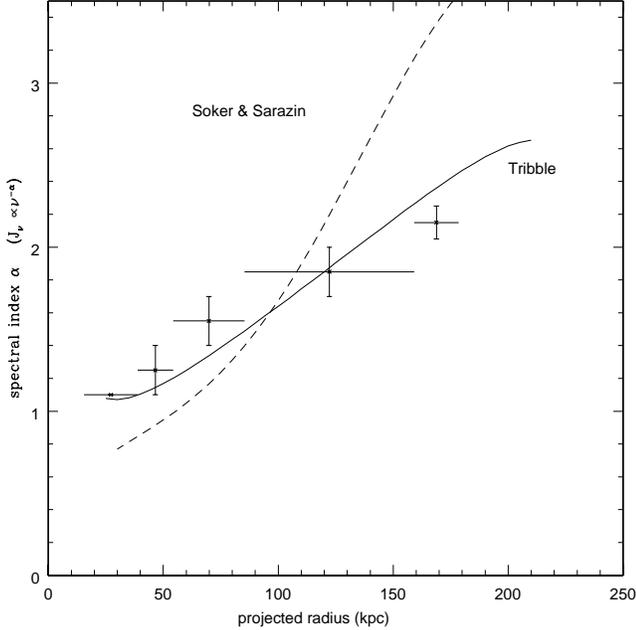}
\vspace{8cm}
\caption{\textit{Solid line}: fit to the spectral steepening obtained 
with the same set of parameters of Fig. \ref{fit}.
The data are taken from spectral index distribution map of Sijbring (1993).
\textit{Dashed line}: predicted spectral steepening in the case of radial 
field, for the values of parameters which match the observed
brightness profile and total spectrum.}
\label{irr}
\end{figure}

The Perseus cluster (Abell426), at a redshift $z=0.0183$, is one of the 
brightest clusters in the sky in X--ray band and hosts the nearest large 
cooling flow. 
X--ray deprojection analysis of \textit{ROSAT} PSPC surface brightness 
profiles indicates that the mass deposition rate is about 
500 $\mbox{M}_{\odot} \mbox{ yr}^{-1}$ and the cooling radius is about 210 kpc
(Ettori et al. 1998); with these values of $\dot{\mbox{M}}$ and $r_c$,
one estimates $n_c \sim 1.2 \times 10^{-3} \mbox{ cm}^{-3}$.
The dominant galaxy in the cluster core is the active galaxy NGC 1275.
The associated radio source (Perseus A or 3C 84) is one of the brightest 
radio sources (the total flux density at 327 MHz is 40.66 Jy), 
with a complex structure and a mini--halo whose extension $(\sim 15')$ is
comparable with that of the cooling flow region. 
The radio mini--halo has a steep spectrum with spectral index $\alpha = 1.4$
(between 327 MHz and 609 MHz, $F_{\nu} \propto \nu^{-\alpha}$)
and its total flux is 17.57 Jy at 327 MHz; in addition there is evidence for 
a spectral steepening in the outer regions (Sijbring 1993). 
\\
Because of the simultaneous presence of the radio mini--halo and of the cooling
flow, we can apply to the Perseus cluster the model for electron 
reacceleration presented in the previous Section.
For comparison with the model expectations we have considered the 92 cm 
(327 MHz) radio image shown in Fig. \ref{mappa}. 
In this map the bright radio core of 3C 84 can be very well approximated
by a point--like source and we have estimated that the diffuse mini--halo 
emission is affected by this intense central emission only within 
$\sim 1'$ from the cluster center, in agreement with \textit{ROSAT} and 
Chandra imaging which show an interaction between the radio lobes of 3C 84 and 
the ICM only on this scale (B\"oringer et al. 1993; Fabian et al. 2000). 
For this reason, our model should be applicable for $r$ \gtsim $30$~ kpc.  
\\  
Fig. \ref{mappa} shows that the diffuse mini--halo is 
rather asymmetric so that, because of the spherical symmetry assumed in our 
model, in order to compare the expected and observed radio
emission we have considered a particular direction across the apparently more 
regular region deprived of field radio sources.  
However, it is worth noticing that a strong morphological correlation has been 
shown to hold between the cooling flow X-ray map and the radio mini--halo 
(Ettori et al. 1998; Sijbring 1993). 
This means that our choice of a specific direction does not appear as an 
effective restriction to the conclusions we are led to by the application of 
our spherically symmmetric model to such an asymmetric object.
On the contrary, our results could reasonably be applied to any 
other direction, once the correct values for the physical 
parameters of the cooling flow in that particular direction are taken into 
account (Ettori et al. 1998).
Moreover, this strong morphological correlation can be taken as a hint to the 
influence of the cooling flow over the relativistic plasma behaviour, 
hence supporting the idea of the electron reaccleration being powered by the 
cooling flow itself.
\\
In order to compare the model with the observations, the observational 
constraints are:\\
\textit{brightness profile}: 
we have calculated the surface brightness profile expected by our model 
by integrating the synchrotron emissivity (Eq. \ref{spettro}) 
along the line of sight;\\
\textit{total spectrum}:
the total synchrotron spectrum is obtained by integrating the synchrotron 
emissivity over the cluster volume taking into account the $x$--dependence of 
the parameters involved in the calculation;\\
\textit{radial spectral steepening}:
at each distance $x$ we obtain from Eq. \ref{spettro} the synchrotron 
emissivity as a function of the frequency measured in terms of the break 
frequency and then compute the spectral index between 327 MHz and 609 MHz.
\\   
We have separately considered the two cases of field compression 
previously discussed (see Sec. \ref{magnetic}). 
\\
The free parameters of our model are the energy density distribution
of the relativistic electrons, i.e. the parameter $s$ of Eq. \ref{norms}, 
and the values at the cooling radius of the magnetic field intensity, 
$B_c$, and of the characteristic distance between two subsequent peaks of 
magnetohydrodynamic turbulence, $l_c$.
In order to investigate the range of parameters required by the model
to match the observations, the calculation has been performed as follows:
for each case of field compression we have fixed $s$ and we have allowed 
the other two parameters $B_c$ and $l_c$ to vary; then, 
for a given set of parameters, we have calculated the brightness profile,
the total spectrum and the radial spectral steepening and compared
them with the observations.
\\
Tribble's case of isotropic compression of the field appears to be
very promising. The best results are obtained for $1.5$ \ltsim $\; s $
\ltsim $ 2.5$.
The parameter space which well reproduces the surface 
brightness profile and the total synchrotron spectrum is in good agreement 
with that producing an acceptable radial spectral steepening. 
In particular, at the $90 \%$ confidence level the values of the parameters
are: $s = 2 \pm 0.25$, $B_c \sim 0.9 \div 1.2 \mu$G, 
$l_c \sim 15 \div 25$ pc,
$l_c$ increasing with increasing the strength of the magnetic field $B_c$.
For these parameters, from Eq. \ref{gammabk}, one obtains that the break
energy at the cooling radius is $\gamma_b \simeq 1500$.
For one set of the parameters which best reproduces all the observational 
contraints we show in Fig. \ref{fit}, \ref{flussi} and \ref{irr}
the fits to the surface brightness profile, total spectrum and 
radial spectral steepening.
\\
For $0$  \ltsim $\; s $ \ltsim $ 1.5$ the space of the parameters 
reproducing the surface brightness 
profile is partially consistent with that producing an acceptable total 
synchrotron spectrum, but the derived spectral steepening is too strong.  
It is worth noticing that for $s$ approaching $0$ 
the parameter limits obtained for the 
surface brightness profile and total synchrotron spectrum are not compatible, 
and, therefore, a constant relativistic energy density inside the cooling flow
does not appear to agree with the observations.
This means that the radial distribution of the number density of the 
electrons before reacceleration may not be constant.
For $s$ \ltsim $ 0 \mbox{ and } s $ \gtsim $ 2.5$ all the observational 
constraints can not be reproduced by the same set of parameters. 
\\
On the other hand, our model results, assuming Soker \& Sarazin's case 
of radial field, are not compatible with the observations.
In fact, we can not reproduce the spectral steepening for the same values of 
the parameters which produce the observed radio surface brightness profile 
and total synchrotron spectrum (Fig. \ref{irr}).
The rapid decrease of the magnetic field towards the outer regions    
in Soker \& Sarazin's case selects the relativistic electrons emitting between
327-609 MHz at progressively higher energies, causing a too rapid steepening 
of the synchrotron spectrum.

\section{Discussion and Conclusions}
\label{discussion} 

We have shown that the main properties of the radio mini--halo in the 
Perseus cluster can be accounted for by the synchrotron radiation from 
relic relativistic electrons in the cluster, which are efficiently 
reaccelerated by MHD Fermi type mechanisms due to the compression of the 
cluster magnetic field in the cooling flow region (see Sec. \ref{results}).
Under the assumption of an efficient MHD turbulence with $\delta B / B \sim 
1$ and spherical symmetry, our model has only 3 free parameters, 
namely the magnetic field intensity
$B_c$ and the turbulence scale $l_c$ at the cooling radius and the
radial dependence of the relativistic electron energy density
(parameterized as a power law of exponent $-s$). 
One important conclusion we have reached is that, 
in our model, an isotropic field compression appears to well 
reproduce the observed surface brightness profile and total 
synchrotron spectrum along with the radial spectral steepening. 
The values of the parameters derived by fitting all the observational 
constraints are:
$s= 2 \pm 0.25$, $B_c \sim 0.9 \div 1.2 \mu$G, 
$l_c \sim 15 \div 25$ pc,
$l_c$ increasing with increasing the strength of the magnetic field.
\\
The fact that we can fit the spectral steepening for the same values of the 
parameters which provide a radio surface brightness 
profile and a total synchrotron spectrum in agreement with the observations 
(see Fig. \ref{irr}) demonstrates the internal consistency of the model.
\\
On the other hand, Soker \& Sarazin's arguments leading to a prevailing
radial field in the cooling flow do not appear applicable to the mini--halo 
in the Perseus cluster, as it is not possible to reproduce the radial spectral
steepening for the same set of parameters reproducing the brightness profile 
and total radio spectrum.
\\
Our results showing the goodness of Tribble's model in reproducing all the 
observational constraints are consistent with the physical conditions 
prevailing in the cooling flow region of the Perseus cluster: 
the turbulent velocity 
($v_T = v_A \sim 60 \mbox{ km s}^{-1}$ at the cooling radius) 
is greater than the mean inflow velocity 
($v_F \sim 20 \mbox{ km s}^{-1}$ at the cooling radius), 
and so we expect an isotropic compression of the field 
(see Sec \ref{magnetic}).
Evidence in favour of Tribble's model in a number of cooling flows is also 
provided by Faraday rotation measurements (Taylor \& Perley 1993; 
Ge \& Owen 1993; Taylor et. al. 1990; Perley \& Taylor 1991).
\\
The range of values obtained for $B_c$ is consistent with the measurements of
magnetic fields diffused in the ICM (see Carilli \& Taylor 2002 and references
therein). 
\\
The energetics associated with the reaccelerated electron population in the 
cooling flow region, $\sim 10^{59}$ erg, is smaller but not far from that 
required by detailed modeling for the electrons in the Coma radio halo 
($\sim 5 \times 10^{59}$ erg, case $c$ with $f=0.02$ in Brunetti et al. 2001). 
The number of relativistic electrons in the cooling flow region is 
$\sim 10^{62}$. 
This is comparable to the number of the electrons in a typical radiogalaxy 
and may suggest an important role of the AGNs (and possibly of the central
AGN) in the injecton of the electron relic population.
\\ 
Concerning the characteristic turbulence distance scale,
a possible check for our model is whether or not the obtained values of $l_c$ 
could be expected by some physical considerations on the turbulence in the ICM.
In particular, assuming that the energy spectrum of Alfv\'en waves is a power
law of the frequency $\omega$, ($\propto \omega^{-\nu}$, $\nu > 1$, 
Lacombe 1977),
we examine the possible value of the minimum 
frequency $\omega_{min}$, i.e. the wavelength 
$L= 2 \pi v_A / \omega_{min} = 2 \pi / k_{min}$ 
which carries most of the turbulent energy. 
Tsytovich (1972) suggests that the minimum frequency of an Alfv\'en wave 
spectrum is the collisional frequency of thermal protons, so that
$k_{min} \sim k_{pp} \equiv 2 \pi \nu_{pp} / v_A$, where $\nu_{pp} = 3 n_p /
T_p^{3/2}$. For typical values of $n_p$ and $T_p$ in clusters of 
galaxies $(n_p \sim 10^{-3} \mbox{ cm}^{-3}, T_p \sim 10^7 \div 10^8 
\mbox{ K})$, we obtain $L \sim 2 \times 10^{1 \div 2}$ pc, consistent with 
the value of $l_c$ obtained by our model. 
\\
In the framework of our model, the possibility of giving rise 
to an observable radio mini--halo in a cooling flow region 
is related to several physical parameters.
We have found that the electron spectrum near the cooling radius 
has a break energy which, for the values of parameters
derived by fitting all the observational constraints, is approximately 
$\gamma_b \sim 1500$. This value corresponds to the break energy of the 
relativistic electrons outside the cooling flow region, where no significant
diffuse radio emission is detected.
As the typical life--time of $\gamma \sim 1500$ electrons
is $\sim 1.6 \times 10^9$ yr, much less than the
Hubble time, to balance the radiative losses of these electrons 
large--scale turbulence in the cluster volume is required.
This scenario is similar to that of the classical cluster radio halos where, 
however, as the average magnetic field strength is considerably smaller than 
that in the cooling flow, higher energy electrons ($\gamma_b \sim 10^4$) 
are required to emit in the radio band so that the turbulence efficiency 
would have to be correspondingly greater.
In addition, in Perseus the turbulence can not be too high in order to avoid 
the disruption of the cooling flow.
Based on these considerations, we suggest that  
the physical conditions of the ICM in a cooling flow cluster, which also shows 
a radio mini--halo, are intermediate between those which 
lead to the formation of ``standard'' extended cluster radio halos 
and those holding in cooling flows without radio halos. 
This could explain  --- at least qualitatively ---
the rarity of radio mini--halos found in cooling flow clusters. 
The energetics of the turbulence plays the main part in discriminating
between these two opposite situations. 
A turbulence with a high enough energy density will be efficient at 
reaccelerating the relativistic electrons: depending on the length scale of 
the turbulence, these electrons can produce a diffuse synchrotron emission 
either in the form of extended radio halo or radio mini--halo. 
On the other hand, if the energy density of the turbulence is too low, 
there will be no efficient reacceleration of the electrons and consequently 
no formation of diffuse radio emission, independently of the turbulence scale.
Here we stress that the Perseus cluster may represent a borderline
case: 
in fact there is evidence for a relatively recent merger event (Schwarz et al.
1992; Sijbring 1993; Dupke \& Bregman 2001), which may have left a 
fossil turbulence still quite energetic, with a typical scale  
just above the limit for activating a radio halo, but not so 
irrelevant to produce no reacceleration at all.  
\\
As a further step, we can set a quantitative constraint 
on the large--scale turbulence imposing that the reacceleration time 
scale well outside the cooling radius be more than 
$4 \div 5 \times 10^8$ yr,
as derived by models on the formation of extended radio halos in clusters 
(Brunetti et al. 2001).
If this condition is verified then an extended radio halo will not be 
formed.
Assuming MHD Alfv\'enic acceleration, we obtain $l_c (\mbox{pc})$ \gtsim  
$\; 10 \; B_c^2 (\mu\mbox{G}) / (n_c/10^{-3}) $,
which represents the lower limit on the large--scale turbulence
in order to avoid the formation of a radio halo.
We notice that the value of the parameters obtained by our model for 
the Perseus cluster are relatively close (within a factor $\sim 2$) to this 
limit.
\\
High resolution X--ray studies with XMM--Newton have 
recently shown that the soft X--ray spectra of several cooling flow 
clusters of galaxies are inconsistent with standard cooling flow models 
as the gas cools down to about 2--3 KeV, but not at lower temperatures 
(Peterson et al. 2001, Kaastra et al. 2001, Tamura et al. 2001, 
Molendi \& Pizzolato 2001).
How to reconcile cooling flow models with these observations is still a
matter of debate (e.g. Fabian et al. 2001 and references therein), but
it should be stressed that the 
compression of the thermal plasma and of the magnetic field within the 
cooling region is well ascertained. 
As a consequence, even if the cooling flow is somewhere close to the
central galaxy stopped, the basic results of our model remain unchanged: 
once the turbulence has been energized by the flow, it takes a time of order 
of a few Gyr for it to be dissipated.
During this time, turbulence continues to reaccelerate the electrons and 
determines the radio properties of the mini--halo.

\begin{acknowledgements}
We woud like to thank Luigina Feretti for useful discussion. In particular, we
would like to thank the referee Vah\'e Petrosian for very helpful suggestions
which improved the presentation of this paper.
This work is partially supported by the Italian Ministry for University and
Research (MURST) under grant 99GITTPGR Prog.Giov.Ricerc. E.F.1999. 
\end{acknowledgements}

\appendix{}


\begin{thebibliography}{}

\bibitem[]{} Baum S.A., O'Dea C.P. 1991, \textit{MNRAS} \textbf{250}, 737
\bibitem[]{} Blasi P., Colafrancesco S. 1999, 
\textit{Astroparticle Physics} \textbf{12}, 169
\bibitem[]{} B\"oringer H., Voges W., Fabian A.C., Edge A.C., Neumann D.N. 
1993, \textit{MNRAS} \textbf{264}, L25
\bibitem[]{} Borovsky J.E., Eilek J.A. 1986, \textit{Ap.J} \textbf{308}, 929
\bibitem[]{} Brunetti G., Setti G., Feretti L., Giovannini G. 2001 
\textit{MNRAS} \textbf{320}, 365
\bibitem[]{} Burns J.O., Sulkanen M.E., Gisler G.R., Perley R.A.
1992, \textit{Ap.J.} \textbf{388}, L49
\bibitem[]{} Carilli C.L., Taylor G.B. 2002, 
\textit{Annu.Rev.Astron.Astrophys.} \textbf{40}, in press
\bibitem[]{} Dennison B. 1980, \textit{Ap.J} \textbf{239}, L93
\bibitem[]{} Dolag K., En{\ss}lin T.A. 2000, \textit{Astron. Astrophys}
\textbf{362}, 151
\bibitem[]{} Dupke R.A., Bregman J.N. 2001, \textit{Ap.J.} \textbf{547}, 705
\bibitem[]{} Edge A.C., Stewart G.C., Fabian A.C. 1992, \textit{MNRAS}
\textbf{258}, 177
\bibitem[]{} Eilek J.A., Hughes P.A. 1991, in \textit{Beams and Jets in
Astrophysics}, ed. P.A. Hughes (Cambridge University Press)
\bibitem[]{} Ettori S., Fabian A.C., White D.A. 1998, \textit{MNRAS} 
\textbf{300}, 837
\bibitem[]{} Fabian A.C., Nulsen P.E.J., Canizares C.R. 1984, \textit{Nature}
\textbf{310}, 733
\bibitem[]{} Fabian A.C. 1994, \textit{Annu.Rev.Astron.Astrophys.} 
\textbf{32}, 277
\bibitem[]{} Fabian A.C., Sanders J.S., Ettori S., Taylor G.B., Allen S.W., 
Crawford C.S., Iwasawa K., Johnstone R.M., Ogle P.M. 2000, \textit{MNRAS} 
\textbf{318}, L65 
\bibitem[]{} Fabian A.C., Mushotzky R.F., Nulsen P.E.J., Peterson J.R. 2001,
\textit{MNRAS} \textbf{321}, L20 
\bibitem[]{} Feretti L. 2000, astro-ph/0006379
\bibitem[]{} Feretti L., Giovannini G. 1996, \textit{IAUS} \textbf{175}, 333
\bibitem[]{} Ge J.P., Owen F.N. 1993, \textit{AJ} \textbf{105}, 778
\bibitem[]{} Ginzburg V.L., Syrovatskii S.I. 1965, 
\textit{Annu.Rev.Astron. Astrophys.}, \textbf{3}, 297
\bibitem[]{} Govoni F., En{\ss}lin T.A., Feretti L., Giovannini G. 2000,
\textit{Astron.Astrophys.} \textbf{369}, 441
\bibitem[]{} Hamilton R.J., Petrosian V. 1992, \textit{Ap.J.} \textbf{398}, 350
\bibitem[]{} Kaastra J.S., Ferrigno C., Tamura T., Paerels F.B.S., Peterson 
J.R., Mittaz J.P.D. 2001, \textit{Astron. Astrophys.} \textbf{365}, L99  
\bibitem[]{} Kardashev N.S. 1962, \textit{Soviet Astronomy -- AJ}
\textbf{6}, 3
\bibitem[]{} Lacombe C. 1977, \textit{Astron. Astrophys.} \textbf{54}, 1
\bibitem[]{} Liang H., Hunstead R.W., Birkinshaw M, Andreani, P. 2000, 
\textit{Ap.J.} \textbf{544}, 686
\bibitem[]{} Melrose D.B. 1980, \textit{Plasma Astrophysics: Nonthermal 
Processes in Diffuse Magnetized Plasmas}, Gordon and Breach
\bibitem[]{} Molendi S., Pizzolato F. 2001, \textit{Ap.J.} \textbf{560}, 194
\bibitem[]{} Owen F.N., Eilek J., Kassim N.E. 2000, \textit{Ap.J} 
\textbf{543}, 611
\bibitem[]{} Perley R.A., Taylor G.B., 1991, \textit{AJ} \textbf{101}, 1623
\bibitem[]{} Peterson J.R., Paerels F.B.S., Kaastra J.S., Arnaud M., Reiprich 
T.H., Fabian A.C., Mushotzky R.F., Jernigan J.G., Sakelliou I. 2001,
\textit{Astron. Astrophys.} \textbf{365}, L104 
\bibitem[]{} Petrosian V. 2001, \textit{Ap.J.} \textbf{557}, 560
\bibitem[]{} Sarazin C.L. 1999, \textit{Ap.J.} \textbf{520}, 529
\bibitem[]{} Sarazin C.L. 2001, astro-ph/0105418, to appear in
\textit{Merging Processes in Clusters of Galaxies}, ed. L. Feretti, 
I. M. Gioia, G. Giovannini (Dordrecht: Kluwer), in press  
\bibitem[]{} Schwarz R.A., Edge A.C., Voges W., Boeringer H., Ebeling H., 
Briel U.G. 1992, \textit{Astron. Astrophys.} \textbf{256}, L11
\bibitem[]{} Sijbring D. 1993, \textit{A Radio Continuum and HI 
Line Study of the Perseus Cluster}, PhD Thesis, Groningen
\bibitem[]{} Soker N., Sarazin C.L. 1990, \textit{Ap.J.} \textbf{348}, 73
\bibitem[]{} Tamura T., Kaastra J.S., Peterson J.R., Paerels F.B.S., Mittaz 
J.P.D., Trudolyubov S.P., Stewart G., Fabian A.C., Mushotzky R.F., Lumb D.H., 
Ikebe Y. 2001, \textit{Astron. Astrophys.} \textbf{365}, L87
\bibitem[]{} Taylor G.B., Perley R.A., Inoue M., Kato Y., Tabara H., Aizu K.,
1990 \textit{ApJ} \textbf{360}, 41 
\bibitem[]{} Taylor G.B., Perley R.A. 1993, \textit{Ap.J.} \textbf{416}, 554
\bibitem[]{} Tribble P.C. 1993, \textit{MNRAS} \textbf{263}, 31
\bibitem[]{} Tsytovich V.N. 1972, \textit{An Introduction to the Theory
of Plasma Turbulence}, Pergamon Press

\end{thebibliography}
\end{document}